\documentclass{article}

\usepackage[a4paper,width=170mm,top=25mm,bottom=25mm]{geometry}

\usepackage{cite}
\usepackage{amsmath,amssymb,amsfonts}
\usepackage{algorithmic}
\usepackage{graphicx}
\usepackage{textcomp}
\usepackage{xcolor}
\usepackage{siunitx}
\usepackage{enumitem}
\usepackage[ruled]{algorithm2e}
\usepackage{subcaption}

\usepackage{nomencl}
\makenomenclature
\renewcommand{\nomgroup}[1]{%
\ifthenelse{\equal{#1}{R}}{\item[\textbf{Roman Symbols}]}{%
\ifthenelse{\equal{#1}{G}}{\item[\textbf{Greek Symbols}]}{%
\ifthenelse{\equal{#1}{A}}{\item[\textbf{Acronyms / Abbreviations}]}{%
\ifthenelse{\equal{#1}{sup}}{\item[\textbf{Superscripts}]}{%
\ifthenelse{\equal{#1}{sub}}{\item[\textbf{Subscripts}]}{%
\ifthenelse{\equal{#1}{M}}{\item[\textbf{Mathematical symbols}]}{%
\ifthenelse{\equal{#1}{O}}{\item[\textbf{Other Symbols}]}
\ifthenelse{\equal{#1}{U}}{\item[\textbf{Units}]}
{}
}
}
}
}
}
}
}

\usepackage{lineno}

\def\BibTeX{{\rm B\kern-.05em{\sc i\kern-.025em b}\kern-.08em
		T\kern-.1667em\lower.7ex\hbox{E}\kern-.125emX}}

\date{}

\begin{document}

\title{Piecewise-linear modelling with automated feature selection for Li-ion battery end of life prognosis\\

}


\author{Samuel~Greenbank\thanks{Battery Intelligence Lab, Department of Engineering, University of Oxford, UK, OX1 3PJ}\and David~A.~Howey \footnotemark[1] \thanks{Corresponding author}}

\maketitle


\begin{abstract}
	The complex nature of lithium-ion battery degradation has led to many machine learning based approaches to health forecasting being proposed in literature. However, machine learning can be computationally intensive while linear approaches are faster but have previously been too inflexible for successful prognosis. Piecewise-linear models, combined with automated feature selection, offer a fast and flexible alternative without being as computationally intensive as machine learning. Here, a piecewise-linear approach to battery health forecasting was compared to a Gaussian process regression tool and found to perform equally for the median case but significantly better at the 95\textsuperscript{th} percentile. The input feature selection process demonstrated the benefit of limiting the correlation between inputs. Further trials found that the piecewise-linear approach was robust to changing input size and availability of training data.  
\end{abstract}

\section*{Keywords}

Feature selection, Piecewise-Linear, Lithium-ion, Degradation, Health, Bayes



\printnomenclature

\section{Introduction}

Using data-driven approaches for modelling lithium-ion battery health degradation has been the focus of a significant amount of recent literature \cite{li2019lifetimereview}. A bulk of that literature has used machine learning approaches to map to a range of targets, from current state of health (SoH), future SoH or remaining useful life (RUL). Machine learning is a fantastic tool but has significant issues. Training a machine learning tool can be computationally challenging, either by scaling poorly or by requiring extensive data quantities. Storing those models can be a limiting factor for real-world systems and they are hard to use for control purposes \cite{li2019lifetimereview}. Linear models are a simpler alternative \cite{xu2018factoringmarkets}. Here, we propose a piecewise-linear model for forecasting capacity without compromising model performance relative to a machine learning tool.

\nomenclature[A]{SoH}{State of health}
\nomenclature[A]{RUL}{Remaining useful life}

Battery degradation is complex due to the range of possible degradation modes and is further complicated when use and cell-to-cell variability must be considered \cite{birkl2017degradationdiagnostics, reniers2019modelsreview, fleischer2014onlineadaptivepart1}. The flexibility inherent in machine learning approaches produces good results when predicting SoH, RUL and knee points. \cite{li2019lifetimereview, lucu2020B_GP_RUL, fermin2020knees, severson2019data1, goebel2008prognosticsmanagement, liu2010RNNRUL, richardson2019gaussianregressionDeltaSoH}. When used alongside machine learning, feature selection based on correlations can act as a source of modelling flexibility that produces very good results \cite{hu2020SoHRUL_selection_gp, greenbank2020SoH_AutoFeatGen, li2021SoHEstimation_FeatSel_Correlation}.

Linear correlations have been found between various stimuli and SoH. Features of incremental capacity curves have been found to vary linearly with capacity loss \cite{li2018IClinearCapLoss, samad2016CapLossICPeakVoltage, bartlett2015SoH_Estimation_Capacity_Composite_Electrodes, weng2013SoHEstimation_IC_Cap, berecibar2016SoHEstimation, li2016SoHEstimation_PhaseTransition, li2021SoHEstimation_FeatSel_Correlation}. Similarly, properties of temperature profiles have demonstrated linear relationships with capacity \cite{wu2018Entropy-induced_dTdt, barre2013Mech_SoHestimation_review}. Other functions with demonstrated linear relationships with health are voltage regions \cite{willenberg2020JellyRollLowSOC, gou2020SOH_estimation_ensemble, mohtat2019Piecewise_lin_volt_model, li2021SoHEstimation_FeatSel_Correlation}, internal stresses \cite{cannarella2014CapLossLinElectrodeExpansion, bartlett2015SoH_Estimation_Capacity_Composite_Electrodes} and combinations of impedances \cite{saha2007integratedbayesianregression}. 

These linear relationships mentioned were found in specific experimental scenarios so cannot be described as universal. But the existence of approximate linear relationships across the range of use cases suggests that there are correlated features to be found which could produce a linear model for lithium-ion battery ageing. The correlation-based selection process which produced good results for machine learning appeared well-suited to finding a requisite set of inputs for a flexible but faster linear model.

However the flexibility of an automated selection step may not be sufficient to accurately map complex degradation profiles. For example, the ``\textit{knee point}" appears in lithium-ion cells undergoing arduous use and manifests as a sudden collapse in health \cite{fermin2020knees, yang2017KneePoint_LithiumPlating, atalay2020theoryofbatteryageing-manyparams, pugalenthi2020PiecewiseInflectionPoint}. Such changes in degradation can result in features moving from linearly varying SoH to non-linear relationships \cite{yang2017KneePoint_LithiumPlating, mandli2019SoH_Prediction_Resistance_Effect}. A linear model, even if flexibly selected, needs to be adaptable to changing degradation rates.

Piecewise-linear models, if used correctly, should be capable of exactly that required adaptability. They have previously been used in state of charge models \cite{alsharif2014piecewisebattery, fleischer2014onlineadaptivepart1, li2011nonlinearadaptiveobserver, meng2019simplifieddynamiclinearSOC, mohtat2019Piecewise_lin_volt_model} and SoH models \cite{zhang2017prognosticstochastic, xu2018factoringmarkets, xiong2019EVsRULPDF}. A common approach is to split models apart according to the stage in a cell life time \cite{fleischer2014onlineadaptivepart1, alsharif2014piecewisebattery, meng2019simplifieddynamiclinearSOC, xiong2019EVsRULPDF}. However cells degrade at different rates, even if identically used \cite{baumhoefer2014cellvar}. Other approaches have split their linear functions according to voltage or state of charge regions \cite{li2011nonlinearadaptiveobserver, farag2017continuouspiecewiselinearrealtime, wen2004onlinemobile}. It has even been possible to use separate linear models according to how a cell is being used \cite{chu2020stochasticCapLossRUL} or to use locally linear regression, where linear models are constructed based on a number of nearest neighbours, to predict capacity loss \cite{yu2020SoH_Estimation_limitedLabeled_LocalLinearRegression}.

Here we proposed to use piecewise-linear regression (PLR) models to forecast battery SoH to end of life. The approach automated the locations of the boundaries between the linear models and the selection of the variable used to split. Then the number of linear models was chosen automatically based on a compromise between complexity and performance. There was a comparison to using a machine learning approach and a number of smaller investigations into how resilient the proposed piecewise-linear approach is to varying modelling conditions.
\nomenclature[A]{PLR}{Piecewise-linear regression}

\section{Data sources}

This work used open-source battery cycling and capacity fade data from two datasets \cite{severson2019data1, attia2020data2}. The datasets were produced for fast-charging experiments and used lithium iron phosphate/graphite 18650 lithium-ion cells, manufactured by A123, all cycled in a temperature chamber set at \SI{30}{\celsius}. Capacity estimates were calculated from the repeated 4C discharge cycles. 

\nomenclature[U]{C}{C-rate}
\nomenclature[U]{Ah}{Ampere hours}

\begin{figure}[h!]
	\centering
	\includegraphics[width=.7\columnwidth]{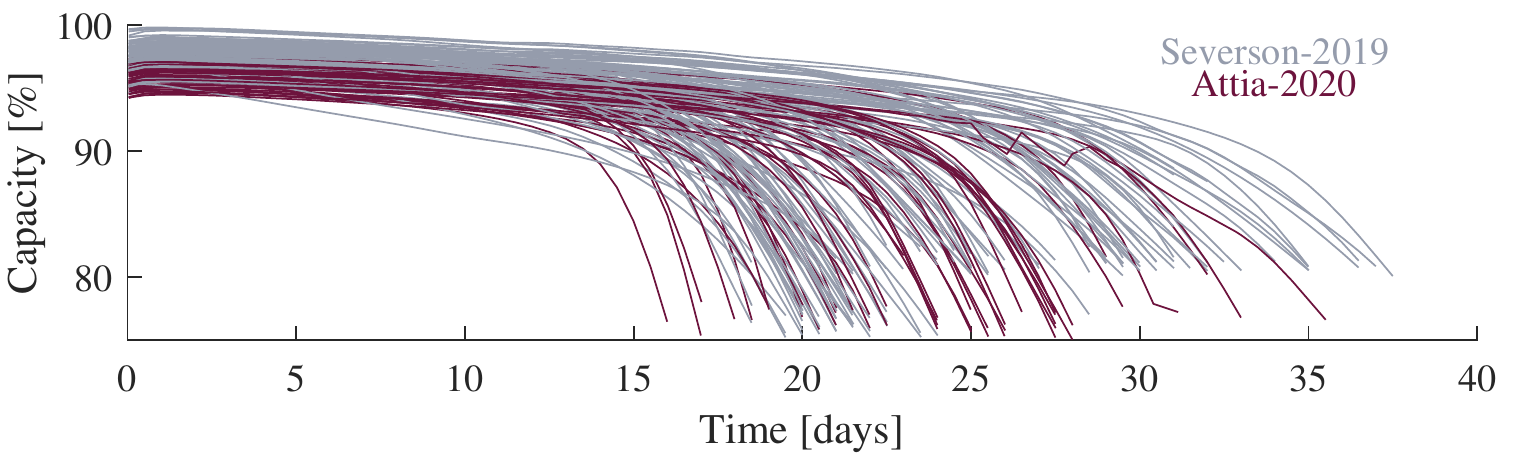}
	\caption{The data used in references \cite{severson2019data1} and \cite{attia2020data2}. Image adapted from ref.\ \cite{greenbank2020SoH_AutoFeatGen}.}
	\label{fig:stanford_capacities}
\end{figure}

Cells were cycled to failure, defined for both as 80\% of the 1.1 Ah nominal capacity. The first work contained 135 cells, with peak charging currents varying between 3.6C and 8C \cite{severson2019data1}. The follow up work, with a further 45 cells, had a fixed charging window of 10 minutes but different paths to full charge \cite{attia2020data2}. Here, all cells with lifetimes between 15 and 40 days were chosen. After this, the data for 157 cells remained available, shown in Fig.\ \ref{fig:stanford_capacities}.

\section{Methods}

\subsection{Data generation and selection} \label{subsec:generation_selection}

The raw data was reduced from 100s of millions of rows down to thousands by producing input features. In each 12 hour time step the input features were calculated based on cell use. The available variables were current, voltage, temperature, power, absolute current and absolute power. 

The input features were all proportions of time spent in given regions of those variables within a specific period in time. The regions were bounded by the behaviour of the full data set. Every variable was split up according to how long was spent at each value so that a cumulative distribution could be produced from which to draw the bounds. 

\begin{table*}
	\begin{center}
		\begin{tabular}{|c|c|c|c|c|c|c|}
			\hline 
			  & Current & Voltage & Temperature & Power & Absolute Current & Absolute Power \\ 
			Percentile & [A] & [V] & [\SI{}{\celsius}] & [W] & [A] & [W] \\ \hline
			1\textsuperscript{st} & -4.00 & 2.00 & 29.8 & -12.8 & 0.00 & 0.0 \\
			33\textsuperscript{rd} & -0.45 & 3.12 & 32.5 & -0.9 & 1.00 & 2.8 \\
			67\textsuperscript{th} & 1.00 & 3.51 & 35.0 & 3.4 & 4.00 & 12.4 \\
			99\textsuperscript{th} & 6.00 & 3.60 & 40.6 & 21.3 & 6.00 & 21.3 \\ \hline
		\end{tabular}
		\vspace{2pt}
		\caption{Variable bounds used to generate features.}\label{tab:variable_bounds}
	\end{center}
\end{table*}

The values for the 33\textsuperscript{rd} percentile represented the value below which the 157 cells cumulatively spend 33\% of their lifetimes. The approach aims to capitalise on past literature where linear relationships have been found between time spent in voltage regions and battery ageing \cite{willenberg2020JellyRollLowSOC, gou2020SOH_estimation_ensemble, mohtat2019Piecewise_lin_volt_model, li2021SoHEstimation_FeatSel_Correlation, greenbank2020SoH_AutoFeatGen}. The thresholds for all variables are shown in table \ref{tab:variable_bounds}.

There were 36 potential input features produced for each cell when calculated for all variables over all variable ranges. Another 36 were added by including how those variables change between time intervals and the final two were experimental time and its square root, leaving 74 features. These values were calculated for each 12 hour interval in a cell's life. The approach will be modelling the changes in capacity $\Delta Q$ over the 12 hour time intervals with capacity estimated from the nearest discharge cycle.

The input features which correlated best with the changes in capacity, $\Delta Q$, were selected in order to reduce that number down to an appropriate set of inputs for a data driven model. Pearson's rank $\rho_P$ was used to evaluate the degree of correlation. Unlike some methods, no selected features were allowed to share a correlation coefficient more than $\rho_{P,max} = 0.85$ so that the input features would cover a wider range of variability. 

\nomenclature[R]{$Q$}{Capacity}
\nomenclature[G]{$\Delta Q$}{Change in capacity}
\nomenclature[G]{$\rho_P$}{Pearson's rank correlation}
\nomenclature[G]{$\rho_{P,max}$}{Maximum shared correlation between input features}

Features were labelled according to the raw variable and the numerical ID of the percentiles which defined the thresholds. For example, the most commonly selected features was V\textsubscript{2,3}, i.e.~the proportion of time spent between the 33\textsuperscript{rd} percentile and the 67\textsuperscript{th}.

\nomenclature[R]{$V_{2,3}$}{Proportion of time spent between voltage percentiles 2 and 3.}

Five input features were used for the main piecewise-linear model here. There is also a small investigation looking at how many features are required, but one who's conclusions are data set specific.

\subsection{Piecewise-Linear splitting}

The splitting process in piecewise-linear modelling searches for the best break points among the training data.

\begin{figure}
     \centering
     \begin{subfigure}[b]{.50\textwidth}
         \centering
         \includegraphics[width=\textwidth]{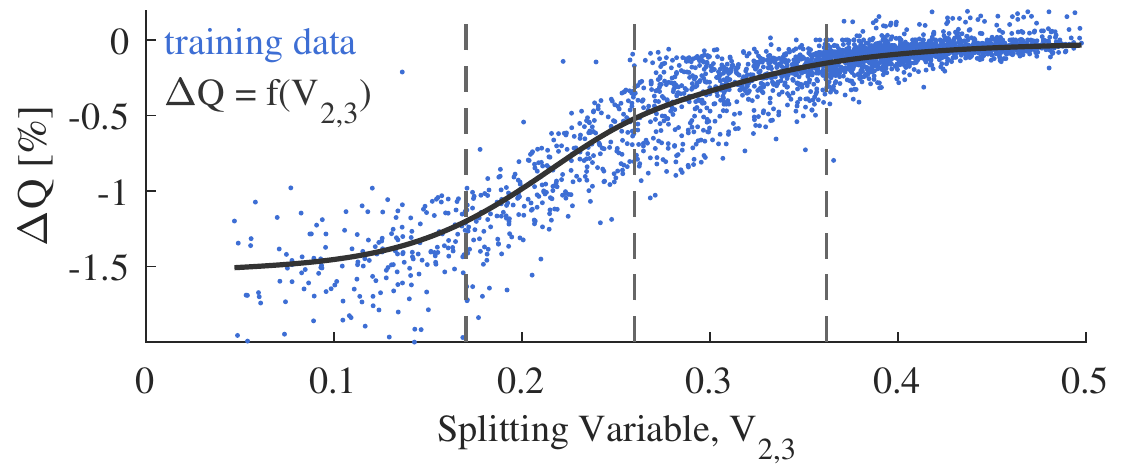}
         \caption{$\Delta Q = \Delta Q(V_{2,3})$.}
         \label{fig:break_point_calc_a}
     \end{subfigure}
     \hfill
     \begin{subfigure}[b]{.45\textwidth}
         \centering
         \includegraphics[width=\textwidth]{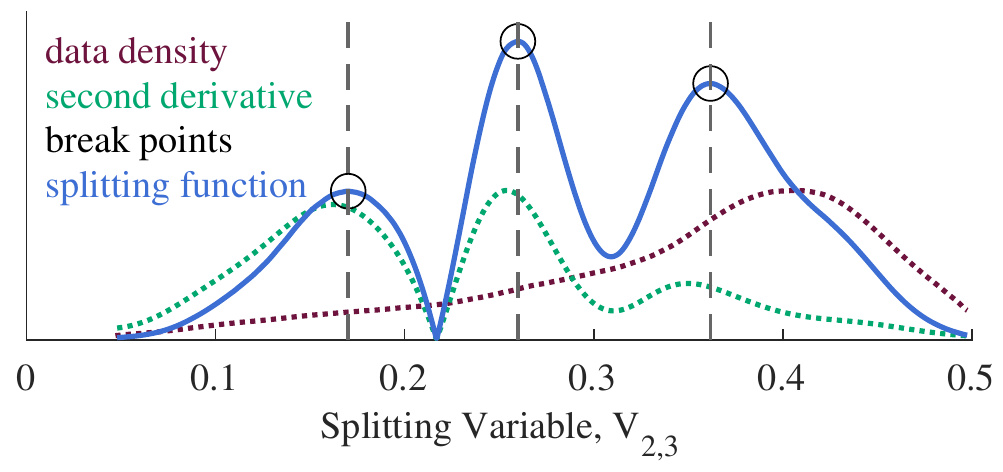}
         \caption{Resultant splitting calculation.}
         \label{fig:break_point_calc_b}
     \end{subfigure}
        \caption{Example calculation of three break points using the best correlating feature, V\textsubscript{2,3}.}\label{fig:example_break_point_calc}
\end{figure}

In all cases, the feature selected first was the input with the best correlation with the loss of capacity. Consequently, that feature was used to calculate the piecewise splits. A function of $\Delta Q = \Delta Q(x)$ was created using a weighted moving average, seen as a black line in Fig.\ \ref{fig:break_point_calc_a}. The second derivative was used to find the points of maximum curvature in Fig.\ \ref{fig:break_point_calc_b}, between which linear models were produced. For clarity, these will be referred to as sub-models from here.

The weights in the moving average were calculated using a squared exponential function of the distance between the target value and the data points. For feature $x$ at value $x_i$, the weight applied to a value of $\Delta Q$ at point $x_j$ was given by:
\begin{equation}
	w_{i,j} = w(x_i,x_j) = \exp{\left(- (x_i-x_j)^2 / \beta_l^2\right)}, \quad \beta_l = \frac{\max(x) - \min(x)}{10}
\end{equation}

\nomenclature[G]{$\beta_l$}{$\Delta Q$ typical lengthscale}
\nomenclature[R]{$w$}{Weighting function}

The function $f_{\Delta Q}(x)$ was then calculated at all points $x_i$ using all $n$ data points in the training set (equation \ref{eqn:weighted_moving_average}).
\begin{equation}\label{eqn:weighted_moving_average}
	f_{\Delta Q}(x_i) = \frac{\sum_{j=1}^{n} w_{i,j} x_j }{ \sum_{j=1}^{n} w_{i,j} }
\end{equation}
\begin{equation} \label{eqn:population_density}
    \rho(x_i) = \frac{1}{n} \sum_{j=1}^{n} \delta \left( ||x_i - x_j|| < \sigma \right)
\end{equation}

\nomenclature[G]{$\rho(x)$}{Data density function}

However, that moving average was likely to have large values for the second derivative at the extreme values of feature $x$ because there were fewer relevant data points. A data density function, $\rho = \rho(x_i)$, was calculated (equation \ref{eqn:population_density}) and multiplied by the second derivative of $f_{\Delta Q}$ so that changing gradients in regions with lots of data points were prioritised. The final expression for the break point selection function, $f_{bp}$, was:
\begin{equation}
	f_{bp}(x_i) = \rho(x_i) \times \left[\frac{d^2 f_{\Delta Q} }{dx^2} \right]_{x=x_i} \nonumber
\end{equation}

\nomenclature[R]{$f_{bp}(x)$}{Break point threshold function}

The process is detailed in Fig.\ \ref{fig:example_break_point_calc}, where the dotted lines in Fig.\ \ref{fig:break_point_calc_b} are multiplied together to produce the final function for selection. Maxima in that composite function matched the significant changes of gradient in the training set or were close by. 

Other approaches to piecewise splitting were considered. K-means and Matlab's \textit{fminsearch} functions were both used as a comparison. K-means was initially used over the full input but that gave too high a weighting to the importance of latterly selected features. K-means was found to be most successful by using just the first two features, at which point the results are extremely similar to that using curvature. Matlab's \textit{fminsearch} allowed for completely free break points selection across the range of the first selected feature. 

\subsection{Linear Regression}

The sub-model for each partition was calculated using Bayesian linear regression. The target variable $y$ was $\Delta Q$ in each time step. It was assumed to be a linear function of input $X$ with associated noise, $\epsilon$. 
\begin{align}
	y = f(X) + \epsilon, \quad \epsilon \sim \mathrm{N}(0,\sigma_n^2) \nonumber
\end{align}
\nomenclature[M]{$y$}{Target variable}
\nomenclature[M]{$X$}{Input array with a data point per row}
\nomenclature[M]{$\epsilon$}{Noise over target variables}
\nomenclature[M]{$\mathrm{N}$}{Normal distribution}
\nomenclature[M]{$\sigma_n$}{Standard deviation of observation noise}
\nomenclature[M]{$f(X)$}{Function of input array, $X$}

Training Bayesian linear regression involves fitting the parameter vector $w$ to create model $f(X)= X w $ based on the posterior distribution over parameters. The input $X$ has a column for every input and a row for every data point.

The parameters $w$ were assigned a mean zero Gaussian prior and covariance $\Sigma_w$.
\begin{align}
	w \sim \mathrm{N}\left(0,\Sigma_w\right) \nonumber
\end{align}
\nomenclature[M]{$w$}{Coefficients of a linear model}

Here, $\Sigma_w$ was assumed to be a diagonal matrix with a constant variance, $\sigma_w^2 = 10^2$. This leads to a mean estimate of the parameters $w$ as a function of the input variables $X$, output target $\Delta Q$ and estimates of observation noise $\sigma_n$ and covariance $\Sigma_w$ \cite{rasmussen2006GPR, murphy2012mathsbook}.
\begin{align}
	\hat{w} = & \sigma_n^{-2} \left( \sigma_n^{-2} X^T X + \Sigma_w^{-1} \right)^{-1} X^T y \label{eqn:blr-full-function} \\
	y_* = & X_* \hat{w} \label{eqn:predicted_cap_loss}
\end{align}
\nomenclature[M]{$\hat{w}$}{Estimate of coefficients}
\nomenclature[sub]{$*$}{Test data}

The predictions of capacity loss $\Delta Q_*$ are therefore produced by then multiplying the test set input matrix, $X_*$, by the parameter estimates in equation \ref{eqn:predicted_cap_loss}.  

\subsection{Piecewise model construction}

The number of sub-models in the final piecewise model $n_m$ was calculated by a compromise between predictive performance and complexity. The procedure is depicted in table \ref{table:selecting_model_size}. All available model sizes are trained on the training set up to some maximum, taken as 10 in the work here. The selected $n_m$ was the minimum model size with an accuracy below the optimal RMSE $\Delta Q$ score multiplied by $(1+\beta_\text{improv})$. For most models here, $\beta_\text{improv} = 0.01$ was chosen.
\nomenclature[R]{$n_m$}{Number of sub-models}
\nomenclature[G]{$\beta_\text{improv}$}{Performance threshold in model selection}

\definecolor{jade}{rgb}{0,.66,.436}
\begin{table*}[h]
	\begin{center}
		\begin{tabular}{|c|c|c|c|}
			\hline 
            $n_m$ & RMSE $\Delta Q$ [\%] & $\leq 1+\beta_\text{improv}$ & selection \\ \hline
            1 & 0.325 &  &  \\
            2 & 0.213 &  &  \\
            3 & 0.201 &  &  \\ \hline
            4 & 0.192 & {\color{jade} 0.192} & $n_m = 4$ \\ \hline
            5 & 0.192 & {\color{jade} 0.192} &  \\
            6 & 0.192 & {\color{jade} 0.192} &  \\
            7 & 0.210 &  &  \\
            8-10 & n/a & & \\
			\hline
		\end{tabular}
		\vspace{2pt}
		\caption{Piecewise model selection by selecting the smallest $n_m$ within $\beta_\text{improv}$ of the peak performance.} \label{table:selecting_model_size}
	\end{center}
\end{table*}

The piecewise-linear model produced $\Delta Q$ estimates over time steps. Capacity profiles were calculated over full cycle life of each test cell by summing the forecasted transitions and assuming knowledge of initial capacity.

\subsection{Performance metrics}

Three performance metrics were calculated for each forecasted capacity profile. Firstly, the root mean squared error (RMSE) observed and predicted $\Delta Q$ provided a measure of performance as a transition model. The capacity profile forecast quality was calculated by using the root mean square error capacity, RMSE Capacity, calculated in \% capacity.
\nomenclature[A]{RMSE $\Delta Q$}{$\Delta Q$ root mean square error}
\nomenclature[A]{RMSE Caapcity}{Capacity root mean square error}

The principle metric used to assess performance was the lifetime accuracy, with end of life (EoL) defined as reaching 80\% nominal capacity. The percentage difference between observed EoL, $\hat{t}_{\text{EoL}}$, and predicted, $t_{\text{EoL}}$, was taken as the error to create the metric PE\textsubscript{EoL}.
\begin{align}
	\text{EoL Error} = 100\% \times ( \hat{t}_\text{EoL}-t_\text{EoL})/\hat{t}_\text{EoL} \nonumber
\end{align}
\nomenclature[A]{EoL}{End of life}
\nomenclature[A]{EoL Error}{End of life percentage error}
\nomenclature[R]{$\hat{t}_\text{EoL}$}{Observed lifetime}
\nomenclature[R]{$t_\text{EoL}$}{Estimated lifetime}

Many trials (minimum 2140 test cells) were performed in at all test points here so that the quoted form of the above metrics will be the median and 95\textsuperscript{th} percentiles.

\subsection{Trial setup}

\begin{figure*}
	\centering
	\includegraphics[width=\textwidth]{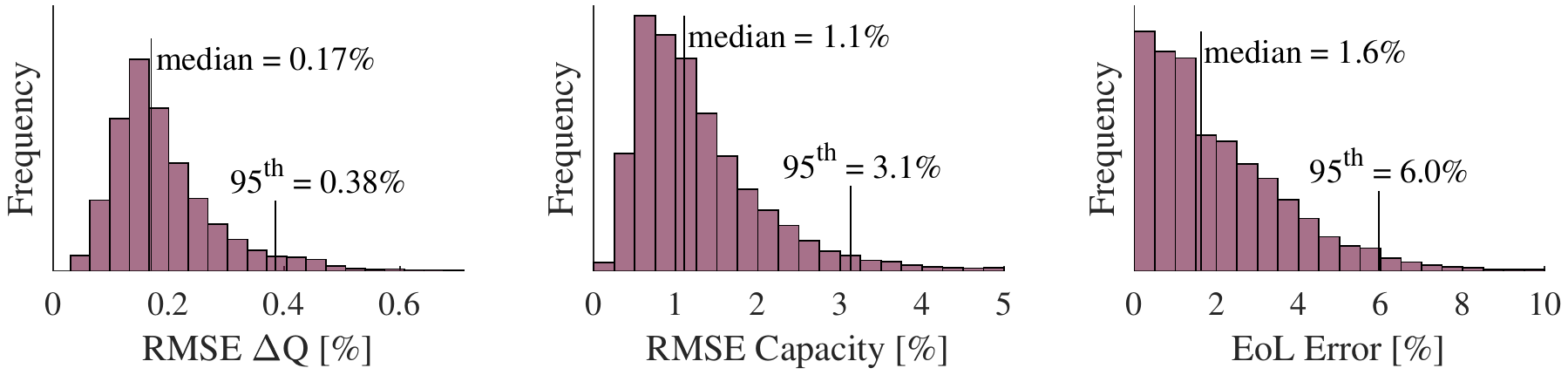}
	\caption{Full results for the piecewise linear modelling. }
	\label{fig:piecewise_main_results}
\end{figure*}

\begin{figure*}
    \centering
    \includegraphics[width=\textwidth]{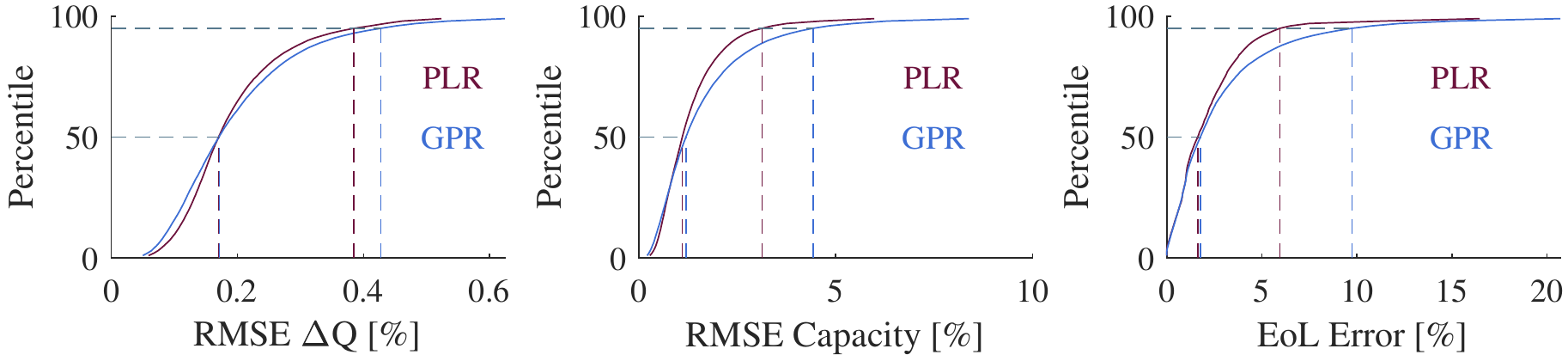}
    \caption{Comparison between piecewise modelling and GPR for capacity forecasting.}
    \label{fig:GPR_comparison_results}
\end{figure*}

The first trial was a comparison in performance against a Gaussian process regression (GPR) tool. The same features were used as the inputs for the piecewise-linear and GPR models so this trial tests the flexibility of using piecewise-linear against that of a machine learning approach. The results for all three performance metrics were calculated for this test. Here, there were 200 repeats of the trial with each repeat using 50 training cells and 107 test cells. Consequently, there are 21,400 predicted capacity profiles to draw from.

The piecewise-linear and automated selection combined approach was also tested to find end of life prediction performance under varying conditions. Smaller tests of end of life error as a function of each of maximum similarity, input features and number of piecewise models were all performed. Each of these held the other parameters from the previous large trial as constant, namely a $\rho_{P,max} = 0.85$, $\beta_\text{improv} = 0.01$, $\max{(n_m)}=10$, 5 input features and 50 training cells. This trial used 20 repeats at each test point each with 107 test cells. 

The last trial calculated the EoL error performance as a function of the number of training cells between 5 and 100. That trial was performed using each of K-means and \textit{fminsearch} as a thorough comparison against the curvature-based piecewise splitting techniques. Apart from the splitting technique, the parameters of this trial was identical to the first one. 

\section{Results}

The distribution of results in Fig.\ \ref{fig:piecewise_main_results} suggested that the piecewise-linear models produced accurate forecasted capacity profiles. The median RMSE $\Delta Q$ of 0.17 \% capacity for the capacity transitions represents a very good fit for the original model. RMSE Capacity was typically very tight with a median value of 1.1 \% capacity. Finally, the predicted lifetimes of the cells were accurate to within 1.6 \% capacity of the observed lifetime in half the test cases. 

The median performance of the piecewise-linear models was extremely similar to that of the GPR models despite GPR performing better at the lowest percentiles. However the 95\textsuperscript{th} percentiles were improved by using piecewise-linear modelling relative to GPR in Fig.\ \ref{fig:GPR_comparison_results}. 

The piecewise-linear approach used between 3 and 5 linear models to map the capacity loss based on the cells here in over 75\% of cases (79\%). In all cases, the input variable used to calculate the thresholds of those linear models was V\textsubscript{2,3} which is the proportion of time spent between 3.12 V and 3.51 V. 

\begin{figure*}
	\centering
	\includegraphics[width=\textwidth]{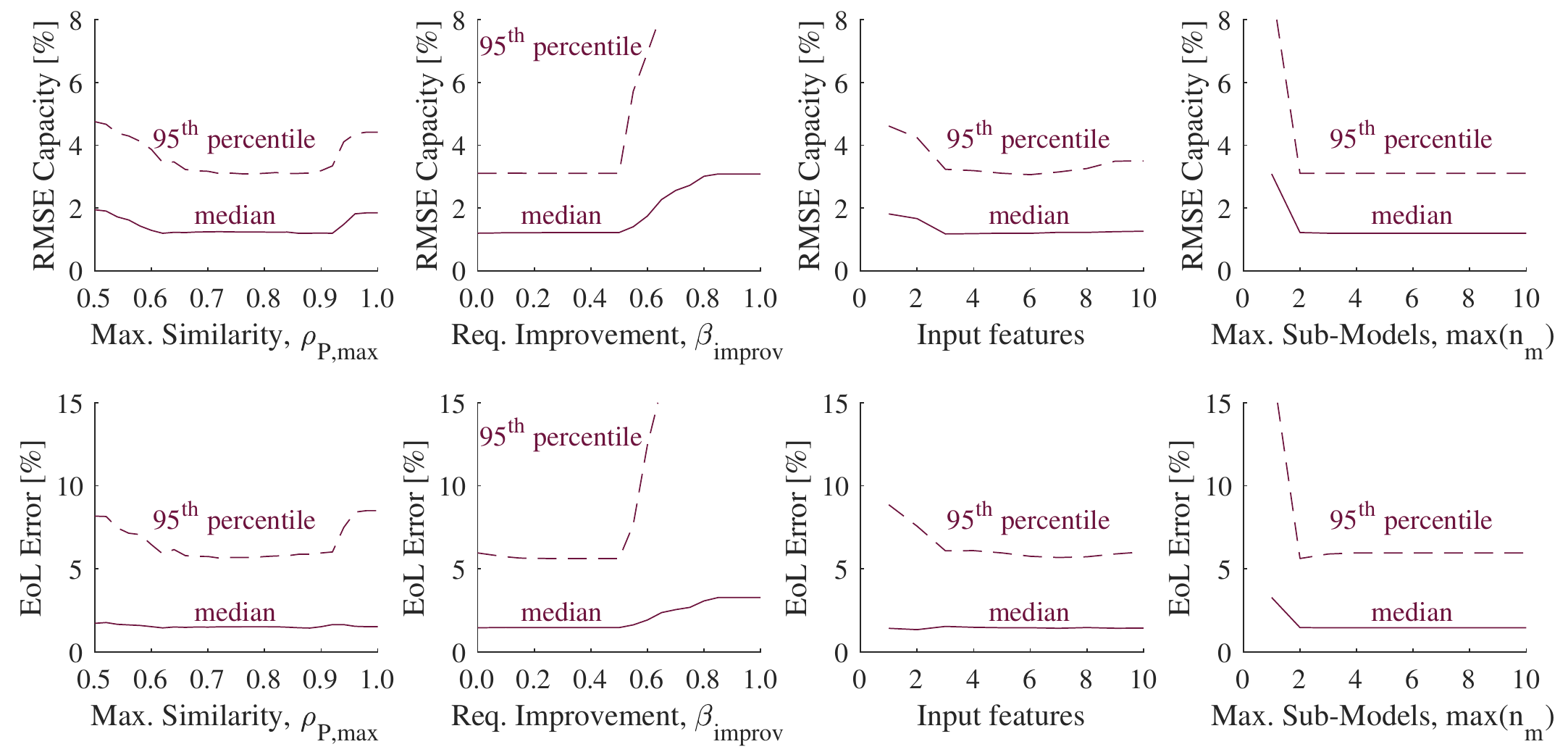}
	\caption{Median and 95\textsuperscript{th} of the end of life predictive performance of the piecewise-linear approach against the maximum correlation among input features, sub-model performance threshold, the number of input features and the maximum number of sub-models.}
	\label{fig:results_maxsim_nfeats_nmods}
\end{figure*}

Median lifetime predictive performance was unaffected over significant changes in control values in Fig.\ \ref{fig:results_maxsim_nfeats_nmods}. Any $n_m$ greater than 1 appeared equally effective, $\beta_\text{improv}$ could be raised to 0.5 without impacting performance and the input feature selection procedure was dependent on $0.6<\rho_{P,max}<0.9$. The 95\textsuperscript{th} percentiles of performance varied more as a function of the control variables. Increasing the number of input features reduced the end of life estimation error up to 3 features, from where the improvements were small before weakening as the number reached 10. 

\begin{figure}
	\centering
	\includegraphics[width=\columnwidth]{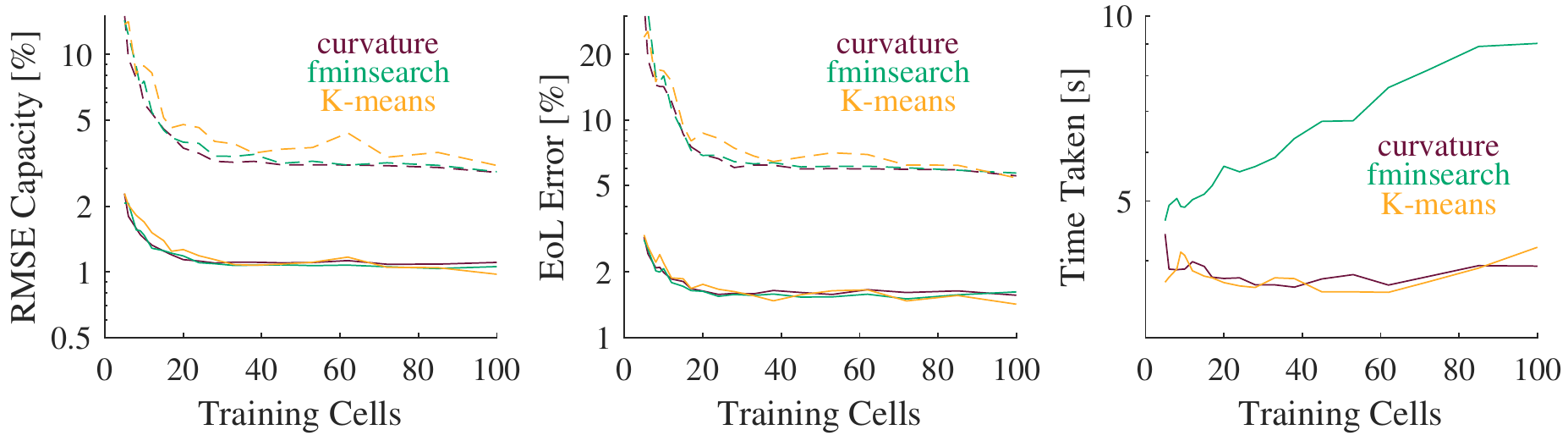}
	\caption{Median and 95\textsuperscript{th} of the performance of the piecewise-linear approach against the number of training cells. Results shown for three splitting techniques: curvature (purple), K-means (yellow) and \textit{fminsearch} (green).}
	\label{fig:results_ntrain_cells_splitting_comp}
\end{figure}

The three splitting methods presented in Fig.\ \ref{fig:results_ntrain_cells_splitting_comp} failed to produce a significant distinction, although calculating using \textit{fminsearch} was found to be far slower with bigger training sets. All three produced good performance with 20 or more training cells. There was only a small improvement by increasing the training set up to 100 cells worth of data.

The histogram in Fig.\ \ref{fig:breakpoints_histogram} shows how there were preferred thresholds for the linear models. The most common was at around V\textsubscript{2,3} $\approx$ 0.37, which roughly corresponded to the change from slow linear ageing to faster degradation in Fig.\ \ref{fig:example_break_point_calc}.

\section{Discussion}

The piecewise-linear approach produced tight capacity forecasts and accurate lifetime prediction. For the same sets of inputs piecewise-linear matched GPR for median performance then outperformed GPR for the 95\textsuperscript{th} percentile of performance. On the other hand, the linear approach appeared less susceptible to smaller patterns in the training data, potentially reducing overfitting. Using over 8 input features produced weaker performance in Fig.\ \ref{fig:results_maxsim_nfeats_nmods}, suggesting that the model was beginning to be overfit.

\begin{figure}[h]
     \centering
     \begin{subfigure}[b]{.31\textwidth}
         \centering
         \includegraphics[width=\textwidth]{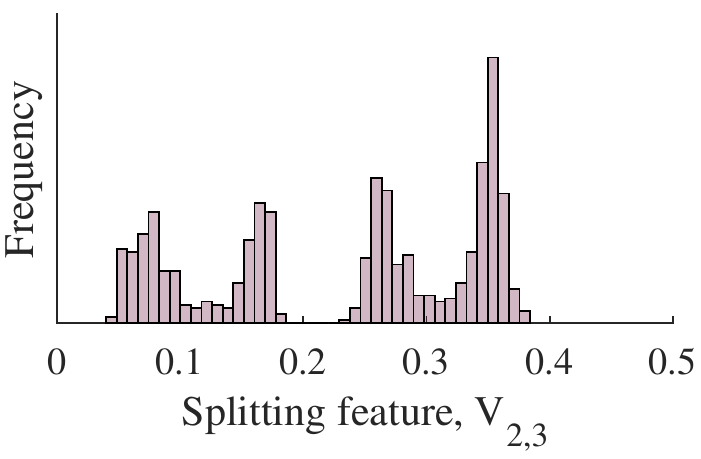}
         \caption{Histogram of selected breakpoints in the large trial in Fig.\ \ref{fig:piecewise_main_results}.}
         \label{fig:breakpoints_histogram}
     \end{subfigure}
     \hfill
     \begin{subfigure}[b]{.31\textwidth}
         \centering
         \includegraphics[width=\textwidth]{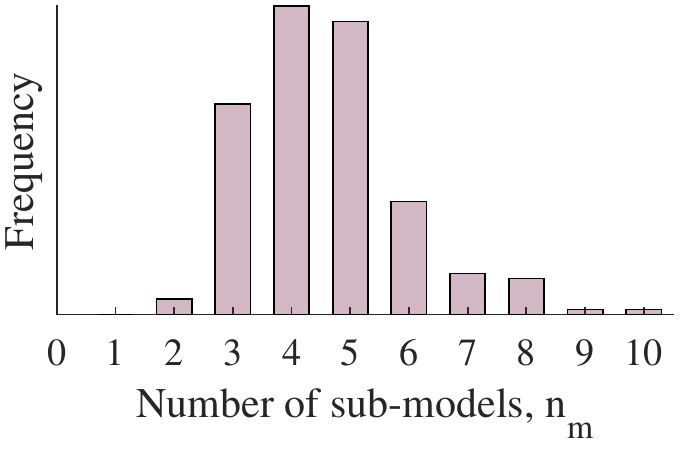}
         \caption{Number of sub-models used in the large trial in Fig.\ \ref{fig:piecewise_main_results}}
         \label{fig:sub_model_count_histogram}
     \end{subfigure}
     \hfill
     \begin{subfigure}[b]{.31\textwidth}
         \centering
         \includegraphics[width=\textwidth]{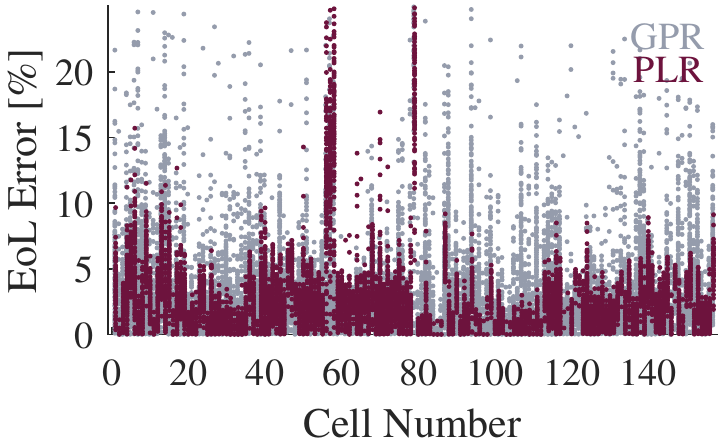}
         \caption{EoL errors for GPR and PLR, sorted by test cell number.}
         \label{fig:eol_error_scatter_comp}
     \end{subfigure}
        \caption{Analysis of the piecewise-linear approach and its results}\label{fig:model_analysis}
\end{figure}

The distinction between the two data-driven approaches appeared to be a function of how poor the poor predictions were. Fig.\ \ref{fig:eol_error_scatter_comp} directly compared the results while grouping the results by test cell. The GPR results for the majority of cells were distributed wider than those for piecewise-linear. The same effect was seen at higher percentiles in Fig.\ \ref{fig:GPR_comparison_results}. Most data-driven approaches struggle to extrapolate, but linear approaches will diverge slower thus reducing the impact of a substandard model. 

With a typical value of 4 linear models used, the piecewise-linear approach was sufficiently detailed to map the more complex trajectories of these rapidly decaying cells. The uniformly primary feature V\textsubscript{2,3} was used to create the boundaries for the linear models. The distribution of those boundaries in Fig.\ \ref{fig:breakpoints_histogram} shows how the region about V\textsubscript{2,3} $\approx$ 0.37 was most popular. That point corresponds to the end of purely linear ageing which is approximately half way through cell life - over 50\% of all data points in the full data set were above that value.

Median performance was consistent across a large range of testing conditions in Fig.\ \ref{fig:results_maxsim_nfeats_nmods}. It must be acknowledged that the quality of the data set contributed to that success. However the selection process guaranteed that input features were chosen based on their linear correlation with the target variable and thus contributed to a model that would work for a majority of cells. 

Successful prognosis for atypical cells required more generous supplies of input data, as demonstrated by the improving 95\textsuperscript{th} percentiles as more cells, inputs, models and input distinction were introduced. There was a distinct improvement when the maximum correlation between inputs was reduced below 0.90, thereby giving more flexibility to the subsequent linear model. That improvement suggested that the maximum shared correlation constraint in the feature selection process was increasing performance.

According to Fig.\ \ref{fig:results_maxsim_nfeats_nmods}, 3 input features and 2 sub-models were required for successful modelling of the majority of the cells in the data set used here. All three piecewise splitting techniques produced good performance from as few as 20 training cells, despite those training cells needing to represent a reasonable range of lifetimes. This efficacy was also a function of the quality of the data set, but it suggested that the piecewise-linear model was capable of good performance even without significant amounts of input data. 

\begin{figure}
    \centering
    \includegraphics[width=.5\columnwidth]{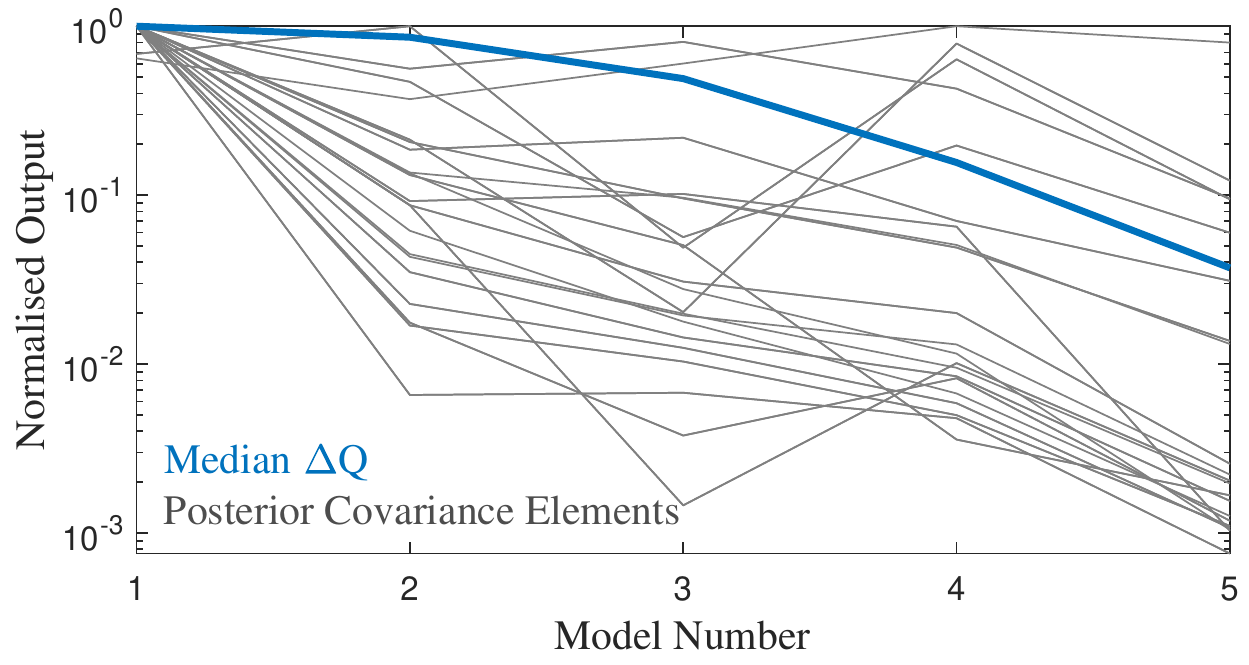}
    \caption{Plot of how posterior covariance elements vary according the piecewise sub-model number. The values are normalised relative to the maximum value found for that item. For comparison, typical degradation rate is also plotted, also normalised relative to the maximum value.}
    \label{fig:posterior_vs_models}
\end{figure}

Detailed evaluation of the performance of the credible intervals produced by Bayesian linear regression is beyond the scope of this work. However the intervals appeared flexible to changing degradation rates, with faster decaying health being associated with increased uncertainty in Fig. \ref{fig:posterior_vs_models}. Each grey line represents the values for a given item in the posterior covariance matrix, equal to $A^{-1}$, within each sub-model. The flexibility afforded by using a piecewise model allowed the posterior covariances to reflect changing uncertainties in addition to the changing degradation rates.

\section{Conclusion}

A combined feature selection and piecewise-linear approach to capacity forecasting was detailed and tested. Under the testing conditions used here, the combined approach produced median RMSE capacity of 1.1\% and median lifetime error of 1.6\%. The piecewise-linear model performed comparably to a Gaussian process regression model when given the same inputs for typical cells, while outperforming the machine learning method at the 95\textsuperscript{th} percentile. The whole approach was robust to reduced input and training set sizes and to limitations being imposed on the piecewise-linear model construction. The feature selection step was shown to improve performance by avoiding input features that correlated too well with each other.

The ability of piecewise-linear models to adapt to wider ranges of use remains uncertain. A user must still be careful to have reliable and appropriate training data. Similarly, uncertainty estimates appeared to be capable of tackling the varied distributions over battery lifetime, but must validated before confident use in the real world.

The work here combines easily understood inputs with simple mechanisms to construct a flexible degradation model that rapidly computes and is easy to store. The model produces accurate capacity forecasts that hold up to and after the collapse of health in later life.


\appendix
\section*{Appendices}

\section{Alternative Splitting Techniques}

The two alternative approaches to finding the break points were using the \textit{fminsearch} Matlab function and using K-means.

The method using \textit{fminsearch} Matlab function represented completely free selection of the break point position. The objective function was RMSE\textsubscript{$\Delta Q$} across the entire training set, and a limit was put in such that breakpoints must be fit in size order.

K-means was performed using the Matlab function \textit{kmeans}. A small trial found that performance was best when using K-means with the first two selected input features, instead of the full training set.

\section{Gaussian Process Regression}

Gaussian process regression (GPR) is a non-parametric, probabilistic approach to regression. It has been used in health and lifetime prediction previously \cite{li2019lifetimereview, lucu2020B_GP_RUL, richardson2019gaussianregressionDeltaSoH, hu2020SoHRUL_selection_gp, greenbank2020SoH_AutoFeatGen}. GPR was used as a comparison to the piecewise-linear model by performing the same mapping between the five automatically selected inputs and the changes in capacity, $\Delta Q$. The rest of the model is identical to the piecewise-linear approach.

The choice of kernel function was a radial basis function (a.k.a. squared exponential), a commonly used stationary covariance function (equation \ref{eqn:kernel_rbf}) \cite{rasmussen2006GPR}. Automatic relevance determination allows for a different length-scale hyperparameter, $ \sigma_{l} $, for each input (equation \ref{eqn:r_ard}). 

\begin{equation}
r(x_i,x_j) = \sqrt{\sum_k \frac{\left(x_{i,k} - x_{j,k}\right)}{\sigma_{l,k}^2}} \label{eqn:r_ard}
\end{equation}
\begin{equation}
\kappa_{rbf}(r) = \sigma_f^2 \exp\left(-r^2\right) \label{eqn:kernel_rbf} 
\end{equation}

The GPR model was the same as the one used in ref.\ \cite{greenbank2020SoH_AutoFeatGen}, but with 50 training cells and the radial basis function kernel.


\begin{thebibliography}{00}
\bibitem{li2019lifetimereview} Y. Li, K. Liu, M. Foley, A. Zulke, M. Berecibar, E. Nanini-Maury, J. Van Mierlo, and H. E. Hoster, ``Data-driven health estimation and lifetime prediction of lithium-ion batteries: a review,'' Renewable and Sustainable Energy Reviews, vol. 113, 2019.
\bibitem{xu2018factoringmarkets} B. Xu, J. Zhao, T. Zheng, E. Litvinov, and D. S. Kirschen, ``Factoring the cycle aging cost of batteries participating in electricity markets," IEEE Transactions on Power Systems, vol. 33, pp 2248--2259, 2018.
\bibitem{fleischer2014onlineadaptivepart1} C. Fleischer, W. Waag, H. Heyn, and D.U. Sauer, ``On-line adaptive battery impedance parameter and state estimation considering physical principles in reduced order equivalent circuit battery models: part 1. Requirements, critical review of methods and modeling," Journal of Power Sources, vol. 260, pp. 276--291, 2014.
\bibitem{birkl2017degradationdiagnostics} C. R. Birkl, M. R. Roberts, E. McTurk, P. G. Bruce, and D. A. Howey, ``Degradation diagnostics for lithium-ion cells,'' Journal of Power Sources, vol. 341, pp. 373-386, 2017.
\bibitem{reniers2019modelsreview} J. Reniers, G. Mulder, and D. Howey, ``Review and Performance Comparison of Mechanical-Chemical Degradation Models for Lithium-Ion Batteries," Journal of the Electrochemical Society, vol. 166, pp. A3189--A3200, 2019.
\bibitem{lucu2020B_GP_RUL} M. Lucu, E. Martinez-Laserna, I. Gandiaga, K. Liu, H. Camblong, W. Widanage, and J. Marco, ``Data-driven nonparametric Li-ion battery ageing model aiming at learning from real operation data - part B: cycling operation," Journal of Energy Storage, vol. 30, 2020.
\bibitem{fermin2020knees} P. Ferm\'in-Cueto, E. McTurk, M. Allerhand, E. Medina-Lopez, M. F. Anjos, J. Sylvester, and G. dos Reis, ``Identification and machine learning prediction of knee-point and knee-onset in capacity degradation curves of lithium-ion cells," Energy and AI, vol. 1, 2020.
\bibitem{severson2019data1} K. A. Severson, P. M. Attia, N. Jin, N. Perkins, B. Jiang, Z. Yang, M. H. Chen, M. Aykol, P. K. Herring, D. Fraggedakis, M. Z. Bazant, S. J. Harris, W. C. Chueh, and R. D. Braatz, ``Data-driven prediction of battery cycle life before capacity degradation,'' Nature Energy, vol. 4, pp. 383--391, 2019.
\bibitem{goebel2008prognosticsmanagement} K. Goebel, B. Saha, A. Saxena, J. Celaya, and J. Christophersen, ``Prognostics in Battery Health Management," IEEE Instrumentation Measurement Magazine, vol. 11, pp. 33--40, 2008.
\bibitem{liu2010RNNRUL} J. Liu, A. Saxena, K. Goebel, B. Saha, and W. Wang, ``An adaptive recurrent neural network for remaining useful life prediction of lithium-ion batteries," Conference of the Prognostics and Health Management Society, 2010.
\bibitem{richardson2019gaussianregressionDeltaSoH} R. Richardson, M. Osborne, and D. Howey, ``Battery health prediction under generalized conditions using a Gaussian process transition model," Journal of Energy Storage, vol. 23, pp. 320--328, 2019.
\bibitem{hu2020SoHRUL_selection_gp} X. Hu, Y. Che, X. Lin, and S. Onori, ``Battery health prediction using fusion-based feature selection and machine learning," IEEE Transactions on Transportation Electrification, 2020.
\bibitem{greenbank2020SoH_AutoFeatGen} S. Greenbank and D. Howey, ``Automated feature selection for data-driven models of rapid battery capacity fade and end of life," pre-print, 2021.
\bibitem{li2021SoHEstimation_FeatSel_Correlation} Y. Li, D. Stroe, Y. Cheng, H. Sheng, X. Sui, and R. Teodorescu, ``On the feature selection for battery state of health estimation based on charging–discharging profiles," Journal of Energy Storage, vol. 33, 2021.
\bibitem{li2018IClinearCapLoss} Y. Li, M. Abdel-Monem, R. Gopalakrishnan, M. Berecibar, E. Maury-Nanini, N. Omar, P. van den Bossche, and J. Van Mierlo, ``A quick on-line state of health estimation method for Li-ion battery with incremental capacity curves processed by Gaussian filter," Journal of Power Sources, vol. 373, pp. 40-53, 2018.
\bibitem{samad2016CapLossICPeakVoltage} N. Samad, Y. Kim, J. Siegel, and A. Stefanopoulou, ``Battery capacity fading estimation using a force-based incremental capacity analysis," Journal of the Elctrochemical Society, vol. 163, pp. A1584--A1594, 2016.
\bibitem{bartlett2015SoH_Estimation_Capacity_Composite_Electrodes} A. Bartlett, J. Marcicki, K. Rhodes, and G. Rizzoni, ``State of health estimation in composite electrode lithium-ion cells," Journal of Power Sources, vol. 284, pp. 642--649, 2015.
\bibitem{weng2013SoHEstimation_IC_Cap} C. Weng, Y. Cui, J. Sun, and H. Peng, ``On-borad state of health monitoring of lithium-ion batteries using incremental capacity analysis with support vector regression," Journal of Power Sources, vol. 235, pp. 36--44, 2013.
\bibitem{berecibar2016SoHEstimation} M. Berecibar, F. Devriendt, M. Dubarry, I. Villarreal, N. Omar, W. Verbeke, and J. Van Mierlo, ``Online state of health estimation on NMC cells based on predictive analytics," Journal of Power Sources, vol. 320, pp. 239--250, 2016.
\bibitem{li2016SoHEstimation_PhaseTransition} X. Li, J. Wang, L. Wang, D. Chen, Y. Zhang, and C. Zhang, ``A capacity model based on charging process for state of health estimation of lithium ion batteries," Applied Energy, vol. 177, pp. 537--543, 2016.
\bibitem{wu2018Entropy-induced_dTdt} Y. Wu, and A. Jossen, ``Entropy-induced temperature variation as a new indicator for state of health estimation of lithium-ion cells," Electrochemica Acta, vol. 276, pp. 370--376, 2018.
\bibitem{barre2013Mech_SoHestimation_review} A. Barr\'e, B. Deguilhem, S. Grolleau, M. G\'erard, F. Suard, D. Riu, S. G\'erard, F. Suard, and D. Riu, ``A review on lithium-ion battery ageing mechanisms and estimations for automotive applications," Journal of Power Sources, vol. 241, pp. 680--689, 2013.
\bibitem{willenberg2020JellyRollLowSOC} L. Willenberg, P. Dechent, G. Fuchs, M. Teuber, M. Eckert, M. Graff, N. K\"urten, D. Uwe Sauer, and E. Figgemeier, ``The development of jelly roll deformation in 18650 lithium-ion batteries at low state of charge," Journal of the Electrochemical Society, vol. 167, 2020.
\bibitem{gou2020SOH_estimation_ensemble} B. Gou, Y. Xu, and X. Feng, ``An ensemble learning-based data-driven method for online state-of-health estimation of lithium-ion batteries," IEEE Transactions on Transportation Electrification, 2020.
\bibitem{mohtat2019Piecewise_lin_volt_model} P. Mohtat, S. Lee, J. Siegel, and A. Stefanopoulou, ``Towards better estimability of electrode specific state of health: decoding the cell expansion," Journal of Power Sources, vol. 427, pp. 101--111, 2019.
\bibitem{cannarella2014CapLossLinElectrodeExpansion} J. Cannarella and C. Arnold, ``State of health and charge measurements in lithium-ion batteries using mechanical stress," Journal of Power Sources, vol. 269, pp. 7-14, 2014.
\bibitem{saha2007integratedbayesianregression} B. Saha, S. Poll, K. Goebel and J. Christophersen, ``An integrated approach to battery health monitoring using bayesian regression and state estimation," IEEE Autotestcon, pp. 646--653, 2007.
\bibitem{yang2017KneePoint_LithiumPlating} X.-G. Yang, Y. Leng, G. Zhang, S. Ge, and C.-Y. Wang, ``Modeling of lithium plating induced aging of lithium-ion batteries: Transition from linear to nonlinear aging," Journal of Power Sources, vol. 360, pp. 28--40, 2017.
\bibitem{atalay2020theoryofbatteryageing-manyparams} S. Atalay, M. Sheikh, A. Mariani, Y. Merla, E. Bower,and W. Widanage, ``Theory of battery ageing in a lithium-ion battery: Capacity fade, nonlinear ageing and lifetime prediction," Journal of Power Sources, vol. 478, 2020.
\bibitem{pugalenthi2020PiecewiseInflectionPoint} K. Pugalenthi, H. Park, and N. Raghavan, ``Piecewise model-based online prognosis of lithium-ion batteries using particle filters," IEEE Access, vol. 8, pp. 153508--153516, 2020.
\bibitem{mandli2019SoH_Prediction_Resistance_Effect} A. Mandli, A. Kaushik, R. Patil, A. Naha, K. Hariharan, A. Kolake, S. Han, and W. Choi, ``Analysis of the effect of resistance increase on the capacity fade of lithium ion batteries," International Journal of Energy Research, vol. 43, pp. 2044--2056, 2019.
\bibitem{alsharif2014piecewisebattery} A. Alsharif, and M. Das, ``A piecewise linear time-varying model for modeling the discharge process of a lithium-ion battery," IEEE International Conference on Electro Information Technology, pp. 587--592, 2014.
\bibitem{li2011nonlinearadaptiveobserver} Y. Li, R. Anderson, J. Song, A. Phillips, and X. Wang, ``A nonlinear adaptive observer approach for state of charge estimation of lithium-ion batteries," Proceedings of the American Control Conference, pp. 370--375, 2011.
\bibitem{meng2019simplifieddynamiclinearSOC} J. Meng, D. Stroe, M. Ricco, G. Luo, and R. Teodorescu, ``A simplified model-based state-of-charge estimation approach for lithium-ion battery with dynamic linear model," IEEE Transactions on Industrial Electronics, vol. 66, pp. 7717--7727, 2019.
\bibitem{zhang2017prognosticstochastic} Z.-X. Zhang, X.-S. Si, C.-H. Hu, and M.G. Pecht, ``A prognostic model for stochastic degrading systems with state recovery: application to Li-ion batteries," IEEE Transactions on Reliability, vol. 66, pp. 1293--1308, 2017.
\bibitem{xiong2019EVsRULPDF} R. Xiong, Y. Zhang, J. Wang, H. He, S. Peng, and M. Pecht, ``Lithium-Ion Battery Health Prognosis Based on a Real Battery Management System Used in Electric Vehicles," IEEE Transactions on Vehicular Technology, vol. 68, pp 4110--4121, 2019.

\bibitem{baumhoefer2014cellvar} T. Baumh\"ofer, M. Br\"uhl, S. Rothgang, and D. U. Sauer, ``Production caused variation in capacity aging trend and correlation to initial cell performance,'' Journal of Power Sources, vol. 247, pp. 332--338, 2014.
\bibitem{wen2004onlinemobile} Y. Wen, R. Wolski, and C. Krintz, ``Online prediction of battery lifetime for embedded and mobile devices," Third Power-Aware Computer Systems, pp. 57-72, 2004.
\bibitem{farag2017continuouspiecewiselinearrealtime} M. Farag, M. Fleckenstein, and S. Habibi, ``Continuous piecewise-linear reduced-order electrochemical model for lithium-ion batteries in real-time applications," Journal of Power Sources, vol. 342, pp. 351--362, 2017.
\bibitem{chu2020stochasticCapLossRUL} A. Chu, A. Allam, A. Cordoba Arenas, G. Rizzoni, and S. Onori, ``Stochastic capacity loss and remaining useful life models for lithium-ion batteries in plug-in hybrid electric vehicles," Journal of Power Sources, vol. 478, 2020.
\bibitem{yu2020SoH_Estimation_limitedLabeled_LocalLinearRegression} J. Yu, J. Yang, Y. Wu, D. Tang, and J. Dai, ``Online state-of-health prediction of lithium-ion batteries with limited labeled data," International Journal of Energy Research, vol. 44, pp. 11345--11360, 2020.
\bibitem{attia2020data2} P. M. Attia, A. Grover, N. Jin, K. A. Severson, T. M Markov, Y-H. Liao, M. H. Chen, B. Cheong, N. Perkins, Z. Yang, P. K. Herring, M. Aykol, S. J. Harris, R. D. Braatz, S. Ermon, and W. C. Chueh, ``Closed-loop optimization of fast-charging protocols for batteries with machine learning,'' Nature, vol. 578, pp. 397--402, 2020.
\bibitem{rasmussen2006GPR} C. E. Rasmussen, and C. K. I. Williams, ``Gaussian processes for machine learning,'' MIT Press, 2006.
\bibitem{murphy2012mathsbook} K. P. Murphy, ``Machine learning: a probabilistic perspective,'' MIT Press, 2012.
\end{thebibliography}
\end{document}